\documentclass[%
 reprint,
superscriptaddress,
 amsmath,amssymb,
 aps,
]{revtex4-1}

\usepackage{graphicx}
\usepackage{dcolumn}
\usepackage{bm}
\usepackage[colorlinks=true,linkcolor=blue,hypertex]{hyperref}%

\begin{document}

\preprint{APS/123-QED}

\title{Electron and ion heating characteristics during magnetic reconnection in MAST}%


\author{H. Tanabe}
\email{tanabe@k.u-tokyo.ac.jp}
\affiliation{Graduate School of Frontier Sciences, University of Tokyo, Tokyo, 113-0032, Japan}
\author{T. Yamada}
\affiliation{Faculty of Arts and Science, Kyusyu University, Fukuoka,  819-0395, Japan}
\author{T. Watanabe}
\affiliation{Graduate School of Frontier Sciences, University of Tokyo, Tokyo, 113-0032, Japan}
\author{K. Gi}
\affiliation{Graduate School of Frontier Sciences, University of Tokyo, Tokyo, 113-0032, Japan}
\author{K. Kadowaki}
\affiliation{Graduate School of Frontier Sciences, University of Tokyo, Tokyo, 113-0032, Japan}
\author{M. Inomoto}
\affiliation{Graduate School of Frontier Sciences, University of Tokyo, Tokyo, 113-0032, Japan}
\author{R. Imazawa}
\affiliation{Japan Atomic Energy Agency, Ibaraki, 311-0193, Japan}
\author{M. Gryaznevich}
\affiliation{Tokamak Solutions, Culham Innovation Centre, Abingdon, OX14 3DB, UK}
\author{C. Michael}
\affiliation{Research School of Physical Sciences and Engineering, Australian National University, Canberra, 0200, Australia}
\author{B. Crowley}
\affiliation{DIII-D National Fusion Facility, General Atomics Court, CA 92186-5608}
\author{N. Conway}
\affiliation{CCFE, Culham Science Centre, Abingdon, Oxfordshire, OX14 3DB, UK}
\author{R. Scannell}
\affiliation{CCFE, Culham Science Centre, Abingdon, Oxfordshire, OX14 3DB, UK}
\author{J. Harrison}
\affiliation{CCFE, Culham Science Centre, Abingdon, Oxfordshire, OX14 3DB, UK}
\author{I. Fitzgerald}
\affiliation{CCFE, Culham Science Centre, Abingdon, Oxfordshire, OX14 3DB, UK}
\author{A. Meakins}
\affiliation{CCFE, Culham Science Centre, Abingdon, Oxfordshire, OX14 3DB, UK}
\author{N. Hawkes}
\affiliation{CCFE, Culham Science Centre, Abingdon, Oxfordshire, OX14 3DB, UK}
\author{the MAST team}
\affiliation{CCFE, Culham Science Centre, Abingdon, Oxfordshire, OX14 3DB, UK}
\author{C. Z. Cheng}
\affiliation{Plasma and Space Science Center, National Cheng Kung University, Tainan, Taiwan}
\author{Y. Ono}
\email{ono@k.u-tokyo.ac.jp}
\affiliation{Graduate School of Frontier Sciences, University of Tokyo, Tokyo, 113-0032, Japan}

\date{\today}

\begin{abstract}
Local electron and ion heating characteristics during merging reconnection startup on the MAST spherical tokamak have been revealed for the first time using a 130 channel YAG-TS system and a new 32 chord ion Doppler tomography diagnostic. 2D local profile measurement of $T_e$, $n_e$ and $T_i$ detect highly localized electron heating at the X point and bulk ion heating downstream. For the push merging experiment under high guide field condition, thick layer of closed flux surface formed by reconnected field sustains the heating profile for more than electron and ion energy relaxation time $\tau^E_{ei}\sim4-10$ms, both heating profiles finally form triple peak structure at the X point and downstream. Toroidal guide field mostly contributes the formation of peaked electron heating profile at the X point. The localized heating increases with higher guide field, while bulk downstream ion heating is unaffected by the change in the guide field under MAST conditions ($B_t>3B_{rec}$).

\begin{description}
\item[PACS numbers]
52.35.Vd, 52.55.Fa, 52.72.+v 
\end{description}
\end{abstract}

\pacs{Valid PACS appear here}
\maketitle


Magnetic reconnection is a fundamental process which converts the magnetic energy of reconnecting fields to kinetic and thermal energy of plasma through the breaking and topological rearrangement of magnetic field lines \cite{Review}\cite{Zweibel}.
   Recent satellite observations of solar flares revealed several important signatures of reconnection heating. In the solar flares, hard X-ray spots appear at loop-tops of coronas together with another two foot-point spots on the photosphere. The loop-top hot spots are considered to be caused by fast shocks formed in the down-stream of reconnection outflow \cite{Masuda_flare}. The two-dimensional (2D) measurements of the “Hinode” spectrometer documented a significant broadening of Ca line-width downstream of reconnection \cite{Hara}. These phenomena strongly suggest direct ion heating by reconnection outflow. On the other hand, the V-shape high electron temperature region was found around X-line of reconnection as an possible evidence of slow shock structure \cite{Shimizu}. However, those heating characteristics of reconnection are still under serious discussion, indicating that direct evidence for the reconnection heating mechanisms should be provided by a proper laboratory experiment. 
Since 1986 the merging of two toroidal plasmas (flux tubes) has been studied in a number of experiments: TS-3 \cite{TS-3_1986}\cite{TS-3_FRC}, START \cite{START_record}, MRX \cite{MRX}, SSX \cite{SSX}, VTF \cite{VTF}, TS-4 \cite{Ono}, UTST \cite{takuma}\cite{Inomoto2015}, and MAST \cite{Mikhail}.
For those laboratory experiments, evidence of plasma acceleration toward outflow direction were observed as split line-integrated distribution function in 0D \cite{SSX_flow}, 1D and 2D bidirectional toroidal acceleration during counter helicity spheromak merging \cite{TS-3oldPRL}\cite{Tanabe_flow}, and in-plane Mach probe measurement  around X point with and without guide field \cite{PPCF}\cite{Jongsoo2014}\cite{Jongsoo}. In the recent TS-3 experiment, with upgrade of diagnostics \cite{OMA}, 2D ion and electron heating characteristics are revealed \cite{PRL} as bulk heating of ions at the downstream region and localized small electron heating around X point.
However, for most of the laboratory experiments, the electron temperature was as low as 15eV due to radiation by low-Z impurities and with the presence of invasive probe diagnostics inside the vessel. The energy inventory has been investigated several times \cite{TS-3_POP}\cite{Hsu}\cite{MRX_nature} but the presence of many loss channels complicated the proper investigation of electron heating and it tends to be underestimated for most of the laboratory experiments.
The world-largest merging device: MAST \cite{Brian} (Mega Ampere Spherical Tokamak) achieved remarkable success in those issues. Although the absence of in-situ probe diagnostics around diffusion region complicates the detailed discussion with magnetics, reconnection heating exceeds $\sim$1keV at maximum \cite{PPCF} both for ions and electrons; in addition, the spatial resolution of Ruby and YAG Thomson scattering diagnostics were recently increased up to 300 and 130 channels respectively \cite{Ruby}\cite{Tom}\cite{Rory}. A new 32 chord tomographic ion Doppler spectroscopy \cite{OMA} measurement has also been installed on midplane with the viewing range around core reconnection region inside $r<0.8$m where the existing charge exchange recombination spectroscopy measurement cannot measure due to the limitation of the impact radii of the neutral beam \cite{Neil}\cite{Ti_CXRS}. This paper addresses the first detailed profile measurement of localized electron and ion heating during magnetic reconnection startup in MAST.
\begin{figure}
\center
 \includegraphics[width=8.6cm]{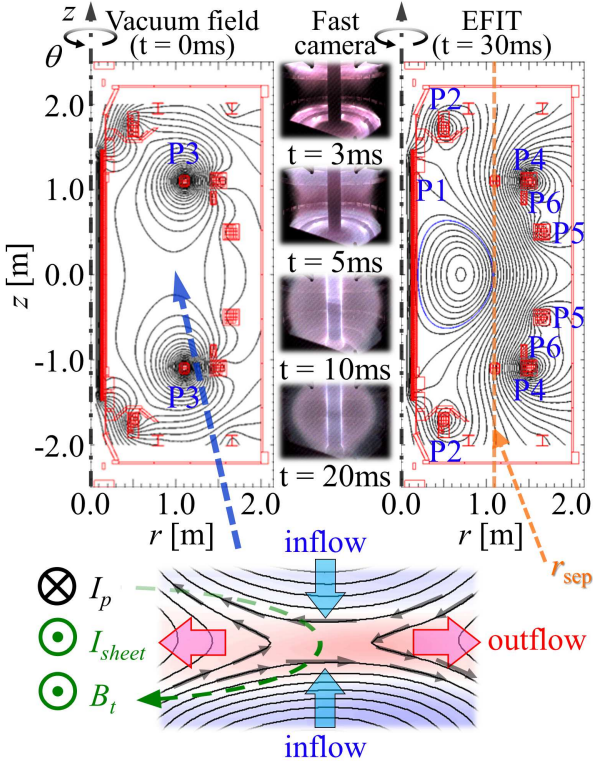} 
 \caption {(color) The geometry of magnetic reconnection in MAST with flux plots  (vacuum field ($t=0$ms) and EFIT ($t=30$ms)) and fast camera images (mostly $\rm D_\alpha$ emission). Two initial STs are generated around P3 coils, move vertically ($z$ direction) and magnetic reconnection is driven at midplane ($z\sim0$m).}
\label{device}
\end{figure}
\begin{figure}
\center
 \includegraphics[width=8.6cm]{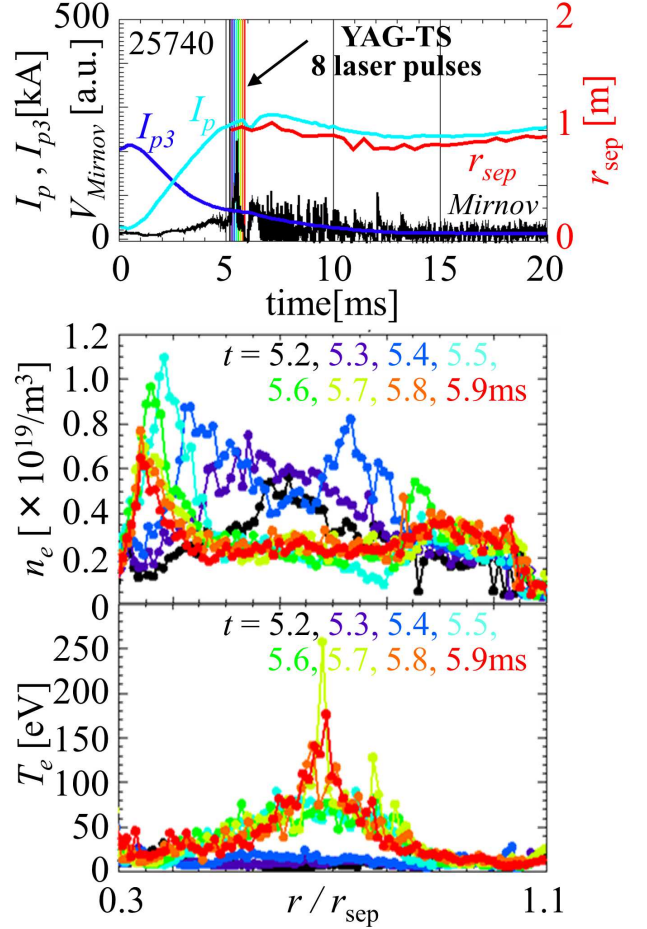}
  \caption{Time resolved Thomson scattering measurement of electron density and temperature profile from 5.2ms with the interval of 0.1ms. During the initial spike of the central Mirnov coil, magnetic reconnection starts and the pile-upped electron density at the X point was ejected towards downstream, while electrons are heated around X-point from $\sim$10eV to $\sim$200eV within 1ms.}
  \label{fig2}
\end{figure}

Figure \ref{device} illustrates the geometry of magnetic reconnection and the definition of the coordinate in MAST. In the cylindrical vacuum vessel ($R_{\rm wall}=2.0$m), P3 coils generate initial two STs which merge together at midplane and mostly contributes to drive magnetic reconnection in MAST \cite{Alan}\cite{Adam_POP}\cite{Ken} (P1 is center solenoid, P2 generates double null divertor configuration after merging, P4 and P5 controls radial equilibrium and P6 coils control the vertical position). 
Toroidal field is $\sim0.3-0.8$T around diffusion region and reconnecting field is roughly $B_{rec}\sim0.07-0.15$T (based on EFIT reconstruction of poloidal $B_r$ field after magnetic reconnection at $t=30$ms: $B_{rec}\sim5.3\times10^{-4}I_{p3\, {\rm max}}{\rm [kA turn]}+2.9\times10^{-3}$); ion skin depth ${\rm c}/\omega_{pi}\sim0.1$m, ion Larmor radius $\rho_{i}<0.01$m and ion cyclotron frequency $\omega_{ci}>10$Mrad/s.
The 300 channel Ruby and 130 channel Nd:YAG Thomson scattering systems measured electron temperature and density at $z=0.015$m and $z=-0.015$m respectively. 32 channel ion Doppler tomography diagnostics measures ion temperature profile at midplane ($z\sim0$m) with spectral resolution of 0.0078nm/pixel for CVI (529.05nm) and has viewing chords in $0.25<r<1.09$m.

 As shown in Fig. \ref{fig2} (top), the P3 ramp down current contributes the formation of initial two STs, magnetic reconnection starts around 5ms with a large spike of central Mirnov coil signal ($\propto dB_z/dt$ at $r\sim0.2$m, $z\sim0$m. The fast camera image in Fig. \ref{device} also shows that two STs move toward midplane around $t=5$ms). During the initial spike of Mirnov coil, 130 channel Thomson scattering measurement of $n_e$ and $T_e$ was performed at 8 time frames in the shot 25740 ($B_{rec}\sim0.11$T and $r_{\rm sep} \sim 1.0$m). Before merging ($t=5.2, 5.3$ms), electron temperature is as low as $\sim10$eV and electron density has a peak around X point. After $t=5.2$ms, magnetic reconnection starts and radial profile of electron density shows clear peak shift which indicates outflow acceleration toward radial direction. For the closed flux type reconnection of spherical tokamak (ST) merging, outflow acceleration is damped at downstream and forms double peak profile with shock-like steep density gradient, while electron temperature rapidly increases at X point with a power density of $\sim 0.3{\rm MW/m^3}$.

\begin{figure}
\center
 \includegraphics[width=8.6cm]{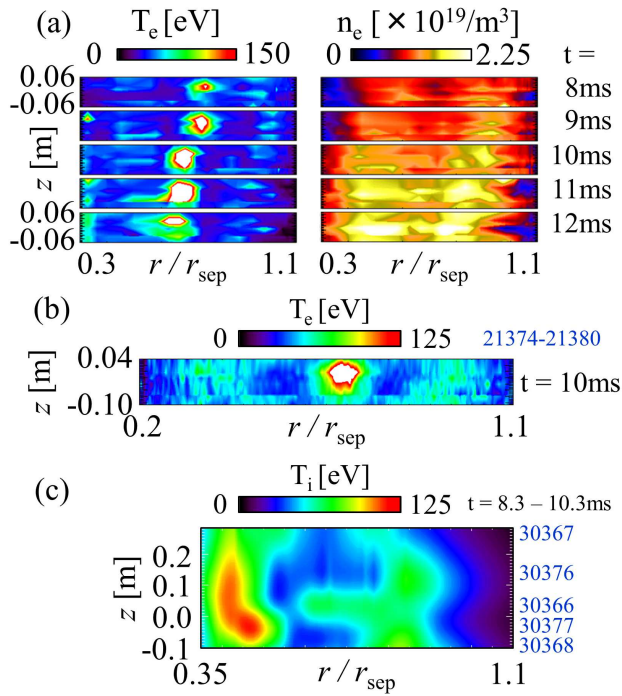}
  \caption{2D Thomson scattering measurement of electron temperature and density profile at t = 8, 9, 10, 11, 12ms around X point (a), 300 channel fine $T_e$ profile using Ruby Thomson scattering measurement (b) and 2D ion temperature profile using 32 chords ion Doppler Tomography diagnostics (c).}
\label{fig3}
\end{figure}

 Figure \ref{fig3} (a) and (b) show 2D electron temperature and density profiles during discharges 21374-21380 (P6 coils are used to shift the vertical position of Thomson scattering measurement). After 5ms, outflow ejection starts and increases electron density downstream, while X point electron heating reaches maximum around $t=10$ms. 
At $t=10$ms, the 300 channel Ruby Thomson scattering measurement is performed. In addition to the highly localized X point electron heating, electron temperature also increases downstream and forms a triple peak structure. The downstream heating is located around the high density region, suggesting the effect of energy relaxation between electrons and ions downstream (unlike most of the laboratory experiments, ion and electron energy relaxation time $\tau^E_{ie}\sim4-10$ms is comparable to the time scale of magnetic reconnection in MAST). Figure \ref{fig3} (c) illustrates the 2D ion temperature profile in pulses 30366-30368, 30376-30377 ($B_{rec}\sim0.08$T). Ions are mostly heated downstream and the region of outflow acceleration around the X-point lies inside the current sheet width (${\rm c}/\omega_{pi}\sim0.1$m) as in two fluid simulation \cite{Adam}. For the high guide field
reconnection experiment in MAST, the ratio of collisional thermal diffusivities $\chi_{\parallel}/\chi_\perp
\sim 2(\omega_{ci}\tau_{ii})^2>>10$ is much higher than that of other laboratory experiments ($\chi_{\parallel}/\chi_\perp \sim1$ for null-helicity operation in MRX \cite{Kuritsyn}). Outflow heating profile downstream is thus confined in a local closed flux surface and enhances the local energy relaxation between ions and electrons, and finally the electron temperature profile also forms peaks at the outflow region.

\begin{figure}
\center
 \includegraphics[width=8.6cm]{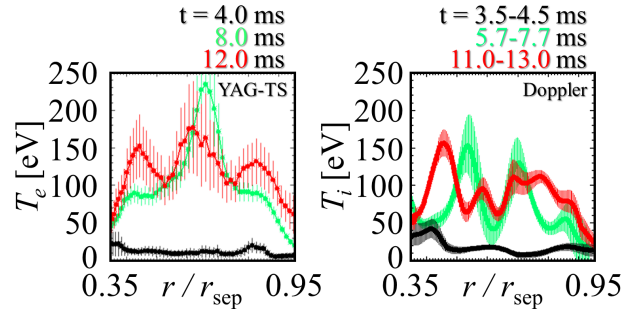}
  \caption{1D radial profile of $T_e$ and $T_i$ at midplane (CIII(464.7nm) line was used in the frame of $t=3.5-4.5$ms). Electrons are mostly heated at X point and ions in the outflow region. Both profiles finally form triple peaks through the energy transfer of ions and electrons with the delay of $\tau^E_{ei}\sim4-10$ms.}
  \label{fig4}
\end{figure}
Figure \ref{fig4} shows more detailed 1D profiles of electron and ion temperature ($B_{rec}\sim0.08$T). Before merging, both temperatures are as low as $\sim10$eV. During magnetic reconnection, electrons are mostly heated at X point, while the ion temperature profile forms double peaks in the outflow region. In MAST experiment, the millisecond time scale of magnetic reconnection is comparable to $\tau^E_{ei}\sim4-10$ms with the result that the localised heating of electrons due to reconnection at the X-point is followed quickly by electron heating in the downstream region due to equilibration with the ions.
Finally both profiles form a triple peak structure at $t\sim12$ms. 

\begin{figure}
\center
 \includegraphics[width=8.6cm]{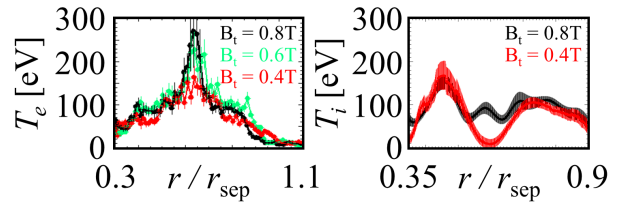}
  \caption{Effect of guide field for electron and ion heating. Under ultra high guide field condition ($B_t > 3B_{rec}$), localized electron heating increases and also enhances ion heating at X point by $T_e - T_i$ energy transfer, while downstream bulk ion heating does not change.}
  \label{fig5}
\end{figure}
Figure \ref{fig5} shows the effect of guide field for electron and ion heating with $B_{rec}\sim0.08$T. The localized X point electron heating becomes more steep and increases under high guide field condition probably because
the higher guide field suppresses cross-field collisional transport, so that the electrons remain in the region of high toroidal electric field for longer
%
or the enhancement of steep sheet current profile for smaller amplitude of meandering motions by higher guide field \cite{Horiuchi}.
However bulk ion heating downstream does not change as demonstrated in the push ST merging experiment with intermittent plasmoid ejection in TS-3 \cite{Plasmoid} and PIC simulation \cite{Inoue}.  
For the operation range of ultra high guide field condition $B_{t} > 0.3$T and $B_t/B_{rec}>3$, outflow dissipation by viscosity damping is suppressed \cite{Braginskii}, however the improved confinement by higher guide field assists the confinement time of ions at downstream in a local closed flux surface,
finally damps outflow and heat ions in the downstream closed flux surface. In addition, for higher guide field reconnection, ion temperature also increases at the X point where the electron temperature profile finally forms triple peak structure.

In summary, localized electron and ion heating characteristics during merging reconnection startup of ST in MAST have been revealed in detail for the first time using 130 channel YAG- and 300 channel Ruby-Thomson scattering measurement and a new 32 chord ion Doppler tomography diagnostic. The 2D profile of electron temperature forms highly localized heating structure at X point
with the characteristic scale length of ${\rm c}/\omega_{ce}< {\rm 0.02-0.05m} <{\rm c}/\omega_{pi}$
, while ion temperature increases inside the acceleration channel of reconnection outflow with the width of ${\rm c}/\omega_{pi} \sim 0.1$m and the downstream where reconnected field forms thick layer of closed flux surface. The effect of $T_i$ - $T_e$ energy relaxation, which has been neglected so far in short pulse laboratory experiments ($\tau_{rec}<<\tau^E_{ei}$), also affects both heating profiles during millisecond timescale reconnection event in MAST. With the delay of $\tau^E_{ei}$ after the maximum heating of electrons at the X point and ions downstream, finally the formation of triple peak structure for both profiles was observed for the first time.
The toroidal guide field mostly contributes to the formation of a localized electron heating structure at the X point and not to bulk ion heating at downstream during merging-type reconnection experiment.
Although the absence of direct magnetic field measurement during reconnection complicated further discussion of the formation mechanism of the characteristics heating profile, it should be noted that the MAST experiment revealed the formation of highly localized heating structure of magnetic reconnection (especially electron heating at X point)  with much improved energy confinement for the first time (in MRX, electron heat loss is comparable to the total energy gain by reconnection \cite{Jongsoo_POP}). In addition, the achieved bulk electron heating reaches comparable order to ion heating after the delay of $\tau^E_{ei}$ and succeeded in pioneering the application of reconnection heating for CS-less startup of spherical tokamak even in the ultra high guide field regime ($B_t > 0.3$T) which is preferable for better confinement in practical operation.

This work was supported by Grants-in-Aid for Scientific Research (A), No 22246119 and Grant-in-Aid for Challenging Exploratory Research NO 2265208), JSPS Core-to-Core program No 22001, the JSPS institutional Program for Young Reearcher Overseas Visists and NIFS Collaboration Research Programs (NIFS11KNWS001, NIFSKLEH024, NIFS11KUTR060). This work was also partly funded by the RCUK Energy Programme under grant EP/I501045 and the European Communities. The views and opinions expressed herein do not necessarily reflect those of the European Commission. This work was carried out within the framework of the European Fusion Development Agreement. We acknowledge Adam Stanier and Alan Sykes for useful discussion  and Samuli Saarelma, Ian Chapman and Brian Lloyd to manage the campaign shots for the reconnection studies.

\bibliography{your-bib-file}

\end{document}